\documentclass[preprint,12pt]{elsarticle}

\usepackage[T1]{fontenc}
\usepackage{lmodern}
\usepackage{amsmath,amssymb,amsthm}
\usepackage{microtype}
\microtypesetup{expansion=false}
\usepackage[hidelinks]{hyperref}
\usepackage{mathtools}
\usepackage{pgfplots}
\usepackage{subcaption}
\usepackage{float}
\pgfplotsset{compat=1.18}
\journal{Economics Letters}
\biboptions{authoryear,round}

\newtheorem{proposition}{Proposition}
\newtheorem{corollary}{Corollary}

\usepackage{eso-pic}
\AddToShipoutPictureFG*{%
  \AtPageUpperLeft{%
    \put(0,-48){%
      \makebox[\paperwidth][c]{%
        \parbox[t]{0.90\paperwidth}{%
          \centering\normalfont\fontsize{9}{11}\selectfont
          The version of record is published in \textit{Economics Letters}.\\[2pt]
          \url{https://doi.org/10.1016/j.econlet.2026.113234}\\[3pt]
          \textcopyright\ 2026. This manuscript version is made available under the CC-BY-NC-ND 4.0 license.\\
          \url{https://creativecommons.org/licenses/by-nc-nd/4.0/}%
        }%
      }%
    }%
  }%
}

\begin{document}

\begin{frontmatter}

\title{When Trust Attracts Fraud: AI and Trust Arbitrage}

\author{Xieyu Yin}
\ead{yinxieyu@csu.edu.cn}
\author{Fenghua Wen\corref{cor1}}
\ead{wfh@amss.ac.cn}
\cortext[cor1]{Corresponding author.}
\address{Business School, Central South University, Changsha 410083, China}

\begin{abstract}
Trust can attract fraud when it delays verification. We develop a two-market signaling model in which generative AI lowers fabrication, verification, and targeting costs. When fabrication becomes profitable before verification, claim credibility first falls and later recovers. Across markets, higher prior quality can delay verification, creating an interval in which only the lower-quality market checks. If targeting becomes profitable in this interval, deceptive sellers enter the higher-quality but less vigilant market, and their entry can initially reverse its reliability advantage. The inflow also triggers verification and deters further entry. We call this self-limiting mechanism \emph{trust arbitrage}. In the age of generative AI, trust can thus create an endogenous but temporary protection gap that redirects deception across markets.
\end{abstract}


\end{frontmatter}

\section{Introduction}

Trust supports exchange partly by economizing on verification. This benefit can become a vulnerability when deceptive sellers move across markets: a market that checks less may become a more attractive target. Generative AI sharpens this tension by lowering fabrication costs \citep{gans2024,galdin2025} and scaling personalized targeting \citep{simchon2024}. Although generative AI may also improve checking, verification can remain a bottleneck \citep{catalini2026}. In a high-trust market, deception is initially rare, so verification may not yet be privately worthwhile even as fabrication and targeting become cheaper. This protection gap raises our central question: when can trust attract rather than deter fraud?

We answer this question with a two-market signaling model. In the single-market benchmark, if fabrication becomes profitable before verification becomes worthwhile, claim credibility first falls when fabrication begins and later recovers when checking starts. We call this fall-and-recovery pattern a \emph{trust valley}. If verification becomes worthwhile no later than fabrication becomes profitable, the unchecked-fabrication region is empty. Because fabricated claims are rarer in a higher-quality market, higher prior quality can delay verification.

This verification delay creates a cross-market opportunity. If targeting becomes profitable after the lower-quality market starts verifying but before the higher-quality market does, deceptive sellers enter the higher-quality but less vigilant market. If that market remains more reliable when entry begins, their growing inflow can initially reverse its advantage. The inflow eventually triggers verification, which rations further entry. We call this self-limiting mechanism \emph{trust arbitrage}.

Our contribution is to show how endogenous verification can redirect deception toward ex ante higher-quality markets. Relative to models of generative AI and communication within a market \citep{gans2024}, our framework allows deceptive sellers to retarget across markets. \citet{bryangans2026} study how anticipated human use---including trust, verification, and further analysis---shapes the value and optimal training of AI predictions; we examine how endogenous verification interacts with strategic fabrication and seller mobility. \citet{quercioli2015} study counterfeit quality and verification across implicit markets; we instead hold claim technology fixed and vary prior market quality, so differences in verification timing can reverse reliability rankings. This prior-dependent timing also extends costly-verification games \citep{ioannidis2022,sadakane2026}. Unlike conventional crime displacement \citep{maheshri2021}, trust arbitrage is self-limiting: the inflow that can initially reverse the ranking also induces verification that limits further entry.

\section{Model}

The benchmark isolates fabrication and verification; Section~\ref{sec:arbitrage} adds targeting and cross-market entry.

A publicly observed state $a\geq0$ indexes AI capability. Nature draws a seller of type $\theta\in\{H,L\}$, with $\Pr(H)=\pi\in(0,1)$. The seller observes $\theta$ privately; the receiver knows only the prior. Trade with type $H$ gives the receiver $u>0$; trade with type $L$ gives $-d<0$. Any seller who trades earns $b>0$. We interpret higher $\pi$ as greater ex ante market trust: a receiver is more willing to accept positive evidence without checking when genuine sellers are more common.

The seller then presents evidence. A type $H$ seller presents genuine positive evidence at zero cost. A type $L$ seller either presents no evidence or fabricates positive evidence that is observationally identical at cost
\begin{equation}
 f(a)>0,\qquad f'(a)<0.
\end{equation}
After observing positive evidence, the receiver can accept the claim without checking, reject it, or verify the evidence at cost
\begin{equation}
 k(a)>0,\qquad k'(a)<0.
\end{equation}
Verification perfectly reveals whether the evidence is genuine: the receiver accepts verified genuine evidence and rejects detected fabrication. The receiver does not trade without positive evidence. Outside options are zero, and fabrication or verification costs are paid when incurred. We study perfect Bayesian equilibria across AI-capability states; $a$ is not a within-game time variable.

Assume
\begin{equation}
 \pi u-(1-\pi)d>0, \label{eq:pooltrade}
\end{equation}
so the receiver accepts unchecked pooled evidence. Let $x$ denote type $L$'s fabrication probability and $v$ the receiver's verification probability. Define $B(\pi)\equiv(1-\pi)d$. Assume that $f$ and $k$ are continuous and strictly decreasing, with $f(0)>b$ and $k(0)>B(\pi)$, and that both converge to zero as $a\to\infty$. These assumptions yield unique thresholds:
\begin{equation}
 f(a_F)=b,\qquad k(a_V(\pi))=B(\pi). \label{eq:thresholds}
\end{equation}
At the two benchmark thresholds and in the absence of mobile sellers, we select no fabrication at $a_F$ and no verification at $a_V(\pi)$. This benchmark selection is not imposed once entry is introduced.

\section{Credibility and verification delay}

The benchmark has three regimes: no fabrication, unchecked fabrication, and active verification. Claim credibility is the posterior probability that positive evidence is genuine,
\[
\tau(a)\equiv\Pr(H\mid\text{positive evidence}).
\]

\begin{proposition}[Single-market equilibrium]
If $a\leq a_F$, then $x=v=0$ and $\tau=1$. If $a_F<a\leq a_V(\pi)$, then $x=1$, $v=0$, and $\tau=\pi$. If $a>\max\{a_F,a_V(\pi)\}$, the interior mixed equilibrium satisfies
\begin{equation}
 x(a,\pi)=\frac{\pi k(a)}{(1-\pi)[d-k(a)]},\qquad
 v(a)=1-\frac{f(a)}{b}, \label{eq:mix}
\end{equation}
and
\begin{equation}
 \tau(a)=1-\frac{k(a)}{d}. \label{eq:trustrecovery}
\end{equation}
\end{proposition}

Two indifference conditions determine the mixed equilibrium. The receiver's indifference between verification and face-value acceptance determines $x$; the low type's indifference between fabrication and no trade determines $v$. A nonempty unchecked-fabrication region requires $a_F<a_V(\pi)$. If $a_V(\pi)\leq a_F$, verification is worthwhile no later than fabrication becomes profitable, so the unchecked-fabrication region is empty.

\begin{figure}[!t]
\centering
\begin{subfigure}[t]{0.48\textwidth}
\centering
\begin{tikzpicture}
\begin{axis}[
    width=\linewidth,height=4.35cm,
    xmin=0,xmax=10,ymin=0.38,ymax=1.05,
    axis lines=left,
    xlabel={AI capability $a$},
    xlabel style={at={(axis description cs:0.5,-0.18)},anchor=north},
    ylabel={Credibility $\tau(a)$},
    xtick=\empty,ytick={0.55,1},yticklabels={$\pi$,$1$},
    tick label style={font=\scriptsize},label style={font=\small},clip=false]
\addplot[black,very thick,domain=0:2.5] {1};
\addplot[black,very thick,domain=2.5:5.5] {0.55};
\addplot[black,very thick,domain=5.5:10,samples=100] {1-0.45*exp(-0.62*(x-5.5))};
\addplot[only marks,black,mark=*,mark size=1.4pt] coordinates {(2.5,1)};
\addplot[only marks,black,mark=o,mark size=1.6pt] coordinates {(2.5,0.55)};
\addplot[black,densely dotted] coordinates {(2.5,0.40) (2.5,1.02)};
\addplot[black,densely dotted] coordinates {(5.5,0.40) (5.5,1.02)};
\node[anchor=north,font=\scriptsize] at (axis description cs:0.25,-0.015) {$a_F$};
\node[anchor=north,font=\scriptsize] at (axis description cs:0.55,-0.015) {$a_V$};
\node[font=\scriptsize] at (axis cs:4.0,0.69) {credibility loss};
\node[font=\scriptsize,align=center] at (axis cs:7.8,0.73) {verification-driven\\recovery};
\end{axis}
\end{tikzpicture}
\caption{Credibility benchmark}\label{fig:panelA}
\end{subfigure}
\hfill
\begin{subfigure}[t]{0.48\textwidth}
\centering
\begin{tikzpicture}
\begin{axis}[
    width=\linewidth,height=4.35cm,
    xmin=0,xmax=10,ymin=0,ymax=0.68,
    axis lines=left,
    xlabel={AI capability $a$},
    xlabel style={at={(axis description cs:0.5,-0.18)},anchor=north},
    ylabel={Fraud inflow into $H$},
    xtick=\empty,ytick=\empty,
    tick label style={font=\scriptsize},label style={font=\small},clip=false]
\addplot[black,very thick,domain=0:2.5] {0};
\addplot[black,very thick,domain=2.5:6,samples=80] {0.13*(x-2.5)};
\addplot[black,very thick,domain=6:9,samples=80] {0.455-0.1517*(x-6)};
\addplot[black,very thick,domain=9:10] {0};
\addplot[black,densely dotted] coordinates {(2.5,0) (2.5,0.63)};
\addplot[black,densely dotted] coordinates {(4.2,0) (4.2,0.22)};
\addplot[black,densely dotted] coordinates {(6,0) (6,0.63)};
\addplot[black,densely dotted] coordinates {(9,0) (9,0.63)};
\node[anchor=north,font=\scriptsize] at (axis description cs:0.25,-0.015) {$a_T$};
\node[anchor=north,font=\scriptsize] at (axis description cs:0.42,-0.015) {$a_R$};
\node[anchor=north,font=\scriptsize] at (axis description cs:0.60,-0.015) {$a_Q$};
\node[anchor=north,font=\scriptsize] at (axis description cs:0.90,-0.015) {$a_V(\pi_H)$};
\node[font=\scriptsize,align=center] at (axis cs:3.65,0.35) {unchecked\\entry};
\node[font=\scriptsize,align=center] at (axis cs:7.55,0.50) {verification\\active};
\draw[<->,thin] (axis cs:4.2,0.12) -- node[above,font=\scriptsize] {$H$ less reliable} (axis cs:6,0.12);
\end{axis}
\end{tikzpicture}
\caption{Targeting and verification}\label{fig:panelB}
\end{subfigure}
\caption{\textbf{Credibility and entry.} Panel A shows the single-market benchmark. Panel B traces a representative interior-entry path: entry begins above $a_T$, the realized integrity ranking first reverses above $a_R$, and entry reaches the verification boundary at $a_Q$. Throughout $a_R<a<a_Q$, market $H$ is less reliable in the realized equilibrium. Entry is necessary for a strict reversal when $a_R<a\leq\min\{a_0,a_Q\}$; if $a_0<a_Q$, it only deepens the integrity gap thereafter. Above $a_Q$, verification rations entry to zero at $a_V(\pi_H)$; entry remains zero thereafter. Curves are schematic.}
\label{fig:main}
\end{figure}
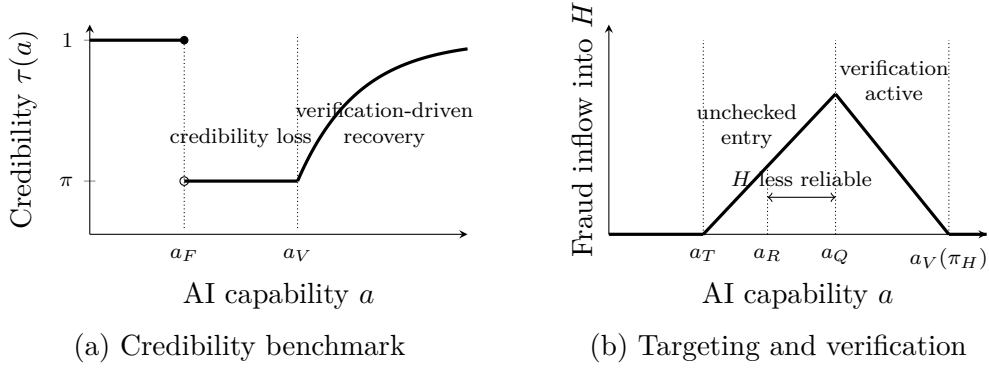

When $a_F<a_V(\pi)$, Proposition 1 gives
\begin{equation}
\tau(a)=
\begin{cases}
1,&a\leq a_F,\\
\pi,&a_F<a\leq a_V(\pi),\\
1-k(a)/d,&a>a_V(\pi)
\end{cases}. \label{eq:valley}
\end{equation}
This nonmonotonic path, shown in Figure~\ref{fig:main}\subref{fig:panelA}, is the trust valley. Claim credibility drops from $1$ to the prior $\pi$ when fabrication starts and recovers only after checking becomes worthwhile. The recovery is strategic: low types continue to fabricate, but cheaper verification reduces their equilibrium fabrication rate.

\begin{proposition}[Verification delay]
The verification threshold increases with prior quality:
\begin{equation}
 \frac{\partial a_V}{\partial\pi}=-\frac{d}{k'(a_V)}>0. \label{eq:delay}
\end{equation}
Hence, when $a_F<a_V(\pi)$, a higher-quality market remains in the pooling region longer. In the active-verification region, $\partial\tau/\partial\pi=0$ and $\partial x/\partial\pi>0$.
\end{proposition}

Fewer low types reduce the expected return to checking pooled evidence. Once verification is active, however, credibility depends on verification cost rather than prior quality. A better prior can then sustain more fabrication while keeping the receiver indifferent.

\section{Trust arbitrage}\label{sec:arbitrage}

We now allow low-type sellers to move across markets. Consider two otherwise identical markets with incumbent priors $0<\pi_L<\pi_H<1$; the labels refer to prior quality. Assume $\pi_Lu-(1-\pi_L)d>0$.

Let $B_j=(1-\pi_j)d$. We focus on states with $a>a_F$ and
\begin{equation}
 B_H<k(a)<B_L. \label{eq:gap}
\end{equation}
In this asymmetric-verification region, the lower-quality market $L$ already verifies while the higher-quality market $H$ still pools. The acceptance condition and $k(a)<B_L$ imply $k(a)<ud/(u+d)$. At the acceptance--verification boundary, the posterior equals $1-k(a)/d$, and the common payoff from unchecked acceptance and verification is $u-k(a)(u+d)/d>0$; hence rejection is strictly worse. By Proposition 1, a low type earns zero expected surplus from fabrication in $L$. A mobile low type must also pay a positive targeting cost and therefore never enters $L$, whereas $H$ can yield positive surplus.

Each market contains an atomless unit mass of incumbent sellers: a mass $\pi_j$ of high types and $1-\pi_j$ of low types. An additional atomless mass $M>0$ of mobile low types can target either market. Claims are randomly matched to an atomless population of receivers. We study symmetric market-level equilibria, so each agent takes aggregate behavior as given.

Each mobile seller draws a targeting friction $\varepsilon$ from a continuous CDF $G$ with support $[0,\bar\varepsilon]$, strictly increasing on that interval. After observing $\varepsilon$, the seller chooses one market or stays out before presenting evidence. The resulting aggregate entry into each market is publicly observed before receivers evaluate evidence. Receivers know that the entrant pool consists entirely of low types, but they cannot identify entrants at the claim level or observe individual seller types.

Targeting either market costs $c(a)+\varepsilon$ in addition to fabrication cost $f(a)$, where $c(a)>0$, $c'(a)<0$, and $c(a)\to0$. The declines in $f$ and $c$ need not be proportional; only the resulting threshold ordering matters. Because entry into $L$ is unprofitable, a seller enters $H$ while it pools if and only if $\varepsilon\leq b-f(a)-c(a)$. Desired entry is the inflow that would occur if $H$ continued to pool:
\begin{equation}
 m^*(a)=M G\!\left(b-f(a)-c(a)\right). \label{eq:mstar}
\end{equation}
The net targeting surplus of a frictionless entrant rises from $-c(a_F)<0$ toward $b>0$. Let $a_T>a_F$ denote its unique zero. Desired entry is zero through $a_T$ and then rises whenever $0<m^*(a)<M$.

The cross-market result requires $a_V(\pi_L)<a_T<a_V(\pi_H)$: targeting becomes profitable after market $L$ starts verifying but before market $H$ does. This threshold ordering restricts the relative cost paths; increasing AI capability alone does not imply it.

Because verification changes which claims result in trade, we compare markets using transaction integrity---the probability of type $H$ conditional on trade---rather than pre-verification claim credibility. While $H$ pools, all incumbent and entrant low types fabricate. After an inflow $m$, market $H$ has transaction integrity $\widetilde\rho_H=\pi_H/(1+m)$. The receiver accepts without checking as long as $\widetilde\rho_H\geq1-k(a)/d$. Hence the largest inflow consistent with unchecked acceptance is
\begin{equation}
 \bar m(a)=\frac{\pi_H}{1-k(a)/d}-1. \label{eq:mbar}
\end{equation}
The lower-quality market $L$ already verifies. Substituting the equilibrium strategies from Proposition 1 gives transaction integrity
\begin{equation}
 \rho_A(a)=\left[1+\frac{k(a)f(a)}{b[d-k(a)]}\right]^{-1}. \label{eq:rhoA}
\end{equation}
\enlargethispage{\baselineskip}
Here the subscript $A$ denotes active verification. Within this region, equilibrium mixing offsets prior quality, which is why $\rho_A$ does not depend directly on $\pi_L$.

To distinguish the realized ranking from its no-entry counterfactual, under the additional condition $\rho_A(a_T)<\pi_H$ used below, let $a_0\in(a_T,a_V(\pi_H))$ be the unique threshold satisfying
\[
 \rho_A(a_0)=\pi_H.
\]
Absent mobile entry, market $H$ has integrity $\pi_H$ and is therefore more reliable than market $L$ precisely when $a<a_0$ within this region.

Define the inflow at which pooling market $H$ becomes less reliable than actively verifying market $L$:
\begin{equation}
 m_R(a)\equiv\max\left\{0,\frac{\pi_H}{\rho_A(a)}-1\right\}. \label{eq:mR}
\end{equation}
Because $\rho_A(a)>1-k(a)/d$, the ranking reverses before entry exhausts $H$'s pooling capacity: $m_R(a)<\bar m(a)$ throughout the asymmetric-verification region. Thus, $H$ can become less reliable while it still accepts claims without checking.

The condition $\rho_A(a_T)<\pi_H$ ensures that the realized ranking first reverses at a strictly positive inflow. Proposition 3 identifies $a_R$, beyond which the ranking is reversed while $H$ still pools, and $a_Q$, where entry reaches $H$'s verification boundary. These conditions are jointly feasible.\footnote{For example, let $u=d=b=M=1$, $(\pi_L,\pi_H)=(0.6,0.8)$, $f(a)=2e^{-a}$, $k(a)=e^{-a}$, $c(a)=e^{-a}$, and let $G$ be uniform on $[0,1]$. Then $a_F=\log 2<a_V(\pi_L)=\log 2.5<a_T=\log 3<a_V(\pi_H)=\log 5$, and $\rho_A(a_T)=3/4<\pi_H$. Here $\rho_A(a)=(1-e^{-a})/(1-e^{-a}+2e^{-2a})$, so $a_0=\log[(1+\sqrt{33})/2]\approx1.216$. Solving the two crossing conditions gives $a_R=\log[(21+\sqrt{273})/12]\approx1.140$ and $a_Q=\log[(25+\sqrt{265})/12]\approx1.235$. Hence $a_T<a_R<a_0<a_Q<a_V(\pi_H)$. Entry is necessary for a strict reversal on $a_R<a\leq a_0$; on $a_0<a<a_Q$, it deepens a ranking that would already be reversed without entry.}

\begin{proposition}[Trust arbitrage]
Suppose $a_V(\pi_L)<a_T<a_V(\pi_H)$ and $\rho_A(a_T)<\pi_H$. There are unique thresholds $a_T<a_R<a_Q<a_V(\pi_H)$ satisfying $m^*(a_R)=m_R(a_R)$ and $m^*(a_Q)=\bar m(a_Q)$. For $a\in(a_R,a_Q)$, market $H$ still pools, but $\rho_A(a)>\widetilde\rho_H(a,m^*(a))$. Moreover, for $a\in(a_R,\min\{a_0,a_Q\})$, $\widetilde\rho_H(a,m^*(a))<\rho_A(a)<\pi_H$, so the reversal would not occur without entry. At $a_Q$, desired entry reaches the pooling boundary. For $a\in(a_Q,a_V(\pi_H))$, pooling is no longer sustainable, so verification is active in $H$.
\end{proposition}

Thus, entry is necessary for the initial ranking reversal and eventually activates verification. If $a_0<a_Q$, then on $(a_0,a_Q)$ entry continues to lower $H$'s integrity, although market $L$ would already be more reliable in the no-entry counterfactual.

Let $G^{-1}$ denote the inverse of $G$ on $[0,1]$.
\begin{corollary}[Post-trigger deterrence]
For $a\in(a_Q,a_V(\pi_H))$, incumbent low types fabricate, and a mobile low type enters and fabricates if and only if $\varepsilon\leq G^{-1}(\bar m(a)/M)$. The symmetric cutoff equilibrium outcome in $H$ is unique and satisfies
\begin{equation}
 m_H(a)=\bar m(a),\qquad
 v_H(a)=1-\frac{f(a)+c(a)+G^{-1}(\bar m(a)/M)}{b}. \label{eq:posttrigger}
\end{equation}
On this interval, $0<\bar m(a)<M$, so the inverse in \eqref{eq:posttrigger} is well defined. Verification rations actual entry at the pooling boundary: $m_H(a)$ strictly decreases and $v_H(a)$ strictly increases. At $a=a_V(\pi_H)$, $x_H=1$, $m_H=0$, and any $v_H\in[1-\{f(a)+c(a)\}/b,\,1-f(a)/b]$, with the functions evaluated at $a_V(\pi_H)$, supports the same zero-entry outcome. For $a>a_V(\pi_H)$, cross-market entry is zero.
\end{corollary}

Delayed verification gives market $H$ a relative entry-surplus advantage for mobile deceptive sellers. Complete exclusion from $L$ follows from four benchmark features: mobility is restricted to low types, receivers know the composition of the entrant pool, verification is perfect, and homogeneous low types earn zero before targeting costs in an actively verifying market. Public observation of aggregate entry allows verification to respond to the realized inflow. Proposition 3 therefore establishes directional displacement under these benchmark conditions.

\newpage
\section{Conclusion}

Generative AI can lower fabrication, verification, and targeting costs at different rates. When fabrication becomes profitable before verification, credibility follows a trust valley. When targeting becomes profitable inside the cross-market verification gap, mobile deceptive sellers enter the higher-quality but less vigilant market. The resulting trust arbitrage is self-limiting: entry can initially reverse the reliability ranking, but it also activates verification and deters further entry. The mechanism concerns the direction of deception, not necessarily its aggregate level. The model therefore identifies conditions under which trusted markets can become temporary targets for fraud in the age of generative AI.

\appendix
\footnotesize
\section*{Appendix: Proofs}

\paragraph{Proof of Proposition 1}
Bayes' rule gives $\mu=\pi/[\pi+(1-\pi)x]\geq\pi$, so \eqref{eq:pooltrade} makes rejection suboptimal on path. If $f\geq b$, the tie break gives $x=v=0$. For $f<b$, pooling with $x=1,v=0$ is supported if and only if $k\geq(1-\pi)d$. Otherwise the receiver's mixing condition is
\[
 \mu u-k=\mu u-(1-\mu)d,
\]
so $1-\mu=k/d$. Bayes' rule gives
\[
 \frac{(1-\pi)x}{\pi+(1-\pi)x}=\frac{k}{d},
\]
which yields $x$ in \eqref{eq:mix}; seller indifference, $(1-v)b=f$, yields $v$. Finally, $\tau=\mu=1-k/d$. \hfill$\square$

\paragraph{Proof of Proposition 2}
Implicit differentiation of $k(a_V)=(1-\pi)d$ gives \eqref{eq:delay}. Equation \eqref{eq:trustrecovery} contains no $\pi$, and from \eqref{eq:mix},
\[
 \frac{\partial x}{\partial\pi}=\frac{k(a)}{[d-k(a)](1-\pi)^2}>0.\qquad\square
\]

\paragraph{Proof of Proposition 3}
While $H$ pools, targeting $H$ yields $b-f(a)-c(a)-\varepsilon$; targeting $L$ yields $-c(a)-\varepsilon<0$. These payoffs imply \eqref{eq:mstar}; desired entry rises whenever $0<m^*(a)<M$. After an inflow $m$, market $H$ has integrity $\pi_H/(1+m)$. Receiver indifference gives \eqref{eq:mbar}, and $k(a)<B_L<ud/(u+d)$ rules out rejection.

Substituting \eqref{eq:mix} into $\Pr(H\mid\text{trade})$ gives \eqref{eq:rhoA}, and
\[
 \rho_A-\left(1-\frac{k}{d}\right)
 =\frac{k(d-k)(b-f)}{d\{b(d-k)+kf\}}>0,
\]
so $m_R<\bar m$. As $f$ and $k$ fall, $\rho_A$ rises, while $m_R$ (when positive) and $\bar m$ fall. At $a_V(\pi_H)$, $k=(1-\pi_H)d$ and $f<b$, so
\[
 \rho_A(a_V(\pi_H))
 =\left[1+\frac{(1-\pi_H)f(a_V(\pi_H))}{b\pi_H}\right]^{-1}
 >\pi_H.
\]
Together with $\rho_A(a_T)<\pi_H$ and the strict increase of $\rho_A$, this establishes existence and uniqueness of $a_0$. Moreover, $m^*-\bar m$, and $m^*-m_R$ while $m_R>0$, are strictly increasing. The condition $\rho_A(a_T)<\pi_H$ gives $m^*(a_T)=0<m_R(a_T)$. As $a\uparrow a_V(\pi_H)$, $\bar m\downarrow0<m^*$, so continuity and strict monotonicity give a unique crossing $a_Q$. At $a_Q$, $m_R<\bar m=m^*$; the same argument gives a unique earlier crossing $a_R$ of $m^*$ and $m_R$. Since $m^*(a_R)=m_R(a_R)>0$, we have $\rho_A(a_R)<\pi_H$ and hence $a_R<a_0$. For $a\in(a_R,a_Q)$, $m_R(a)<m^*(a)<\bar m(a)$, so $H$ pools and $\widetilde\rho_H(a,m^*(a))<\rho_A(a)$. If also $a<a_0$, then $\rho_A(a)<\pi_H$, which proves the counterfactual claim. \hfill$\square$

\paragraph{Proof of the Corollary}
Fix $a\in(a_Q,a_V(\pi_H))$. Since $m^*(a)>\bar m(a)>0$, $v_H=0$ induces excessive entry, whereas $v_H=1$ eliminates entry; the receiver must mix. At the marginal cutoff, a mobile low type is indifferent despite paying $c(a)+\varepsilon>0$. An incumbent low type avoids this cost and strictly prefers to fabricate, so $x_H=1$. The receiver's indifference gives $m=\bar m(a)$. Matching the entry cutoff $\varepsilon\leq(1-v_H)b-f(a)-c(a)$ to this mass yields \eqref{eq:posttrigger} and the unique symmetric cutoff outcome. Since $\bar m<m^*$, the cutoff lies below $b-f-c$, so $0<v_H<1$. After $a_Q$, $f$, $c$, and $G^{-1}(\bar m/M)$ fall, so $v_H$ rises while $\bar m$ falls. At $a=a_V(\pi_H)$, $\bar m=0$, and any $v_H\in[1-\{f(a)+c(a)\}/b,\,1-f(a)/b]$, evaluated at the knife-edge, deters entry while keeping incumbent low types willing to fabricate. For $a>a_V(\pi_H)$, benchmark verification gives every entrant payoff $-c-\varepsilon<0$. \hfill$\square$

\small
\par\bigskip
\noindent\textbf{Funding.}
This work was supported by the National Natural Science Foundation of China [grant number 72131011].

\medskip
\noindent\textbf{Declaration of competing interest.}
The authors declare that they have no known competing financial interests or personal relationships that could have appeared to influence the work reported in this paper.

\medskip
\noindent\textbf{Data availability.}
No data was used for the research described in the article.

\medskip
{\noindent\raggedright\textbf{Declaration of generative AI and AI-assisted technologies in the manuscript preparation process.}\par}
During the preparation of this work, the authors used ChatGPT in order to improve language and readability. After using this tool, the authors reviewed and edited the content as needed and take full responsibility for the content of the publication.

\scriptsize
\setlength{\bibsep}{0pt}

\normalsize

\end{document}